\documentclass[structabstract]{aa}  
\usepackage{graphicx}
\usepackage{txfonts}
\begin{document}
%
% NAMES :
% A-component: HD 16246
% B-component: HD 16232
%
   \title{A substellar component orbiting the F-star 30 Ari B
\thanks{Table 3 is only available in electronic form at the CDS via
  anonymous ftp to cdsarc.u-strasbg.fr (130.79.128.5) or via
  http://cdsweb.u-strasbg.fr/cgi-bin/qcat?J/A+A/; Based on
  observations obtained at the 2-m Alfred-Jensch telescope at the
  Th\"uringer Landessternwarte Tautenburg.  }}

   \author{
          E. W. Guenther\inst{1,2}
          \and
          M. Hartmann\inst{1}
          \and
          M. Esposito\inst{2,3}
          \and
          A. P. Hatzes\inst{1}
          \and
          F. Cusano\inst{1}
          \and
          D. Gandolfi\inst{1}
          }

   \institute{Th\"uringer Landessternwarte Tautenburg,
              Sternwarte 5, D-07778 Tautenburg, Germany\\
              \email{guenther@tls-tautenburg.de}
         \and
                Instituto de Astrof\'\i sica de Canarias,
                C/V\'\i a L\'actea, s/n,
                E38205 -- La Laguna (Tenerife), Spain
         \and
              Hamburger Sternwarte,
              Gojenbergsweg 112,
              D-21029 Hamburg, Germany \\
             }

   \date{Received April 15, 2009; accepted Oct 5, 2009}

% \abstract{}{}{}{}{} 
% 5 {} token are mandatory
 
  \abstract 
% {context} leave it empty if necessary 
{Most current radial velocity planet search programs have concentrated
  on stars of one solar mass. Our knowledge on the frequency of giant
  planets and brown dwarf companions to more massive stars is thus
  rather limited.  In the case of solar-like stars, the frequency of
  short-period brown dwarf companions and very massive planets seems
  to be low.  }
  {Here we present evidence for a substellar companion 
  to 30 Ari B, an F-star of 1.16 $\pm$ 0.04 $\rm M_\odot$
  that is a member of a hierarchical triple system.}
  {The companion was detected by means of precise radial velocity
  measurements using the 2-m Alfred-Jensch telescope and its \'echelle
  spectrograph. An iodine absorption cell provided the wavelength
  reference for precise stellar radial velocity measurements.}
    {We analyzed our radial velocity measurements and derived an orbit
      to the companion with period, $\rm P$ = 335.1 $\pm$ 2.5 days, 
      eccentricity e = 0.289  $\pm$ 0.092, and mass function
      $f(m)$ = (6.1 $\pm$ 1.7) $\times$ 10$^{-7}$ $\rm M_\odot$.
       }
    {We conclude that the radial velocity variations of 30 Ari B are
due to a companion with $m$\,sin\,$i$ of $9.88\pm0.94$ $\rm M_{Jup}$ that is
either a massive planet or a brown dwarf. { The object thus
belongs to the rare class of massive planets and brown dwarfs orbiting
main- sequence stars.}
}


   \maketitle
%
%________________________________________________________________

\section{Introduction}

Although the mass of the host star is likely to be important for
planet formation, our knowledge of the dependence between the mass of
the host star and the frequency and mass of planets is still rather
limited (Udry and Santos (\cite{udry07}) This is, because most planet
search programs focus on dwarfs of about one solar mass, or less.

Although there is strong evidence that stars less massive than the sun
have a lower frequency of giant planets, it still has not been
observationally established whether a different frequency of giant
planets holds for stars more massive than the sun.  A correlation
between the mass of the planets, and the host stars is expected
because the mass of the proto-planetary disk scales with the mass of
the stars (Lovis \& Mayor \cite{lovis07}).  Radial velocity searches
for planets around more massive, early-type main-sequence stars are
problematical due to the paucity of lines and often rapid rates of
rotation -- the RV precision is often inadequate for the detection of
substellar companions. In spite of these difficulties a few searches
have been conducted. Galland et al.  (\cite{galland05a}) surveyed a
sample of A--F stars and discovered a planetary candidate in a 388-day
orbit with $m$\,sin\,$i$ = 9.1 $\rm M_{Jup}$ around the F6V star HD
33564 { which has a mass of $1.25^{+0.03}_{-0.04}$ $M_{\odot}$}
(Galland et al. \cite{galland05b}). They also found a brown dwarf
candidate with $m$\,sin\,$i$ = 25 $\rm M_{Jup}$ in a 28-day orbit
around the A9V star \object{HD 180777} (Galland et
al. \cite{galland06}).

An alternative approach is to conduct RV searches for those early-type
main-sequence stars that have evolved onto the giant branch. The cool
effective temperatures and slow rotation rates make these stars much
more amenable to RV searches even if the mass determination is more
problematical for such stars.  Currently about 20 giant planets have
been found around giant stars in the mass range of 1.1--2.5 $\rm
M_\odot$ (e.g.  Frink et al. \cite{frink02}; Sato et
al. \cite{sato03}; Setiawan et al. \cite{setiawan05} D\"ollinger et
al. \cite {doellinger07}; Lovis \& Mayor \cite{lovis07}; Sato et
al. \cite{sato08}; ).  These planets tend to have masses in the range
of 3--10 $\rm M_{Jup}$ hinting that more massive stars tend to have
more massive planets.  Interestingly, the well known relation between
metallicity and planet frequency for solar-like stars is less
pronounced in giant stars (Schuler et al. \cite{schuler05}).  Pasquini
et al. (\cite{pasquini07}) argued that the planet-metallicity effect
may be a result of atmospheric pollution caused by migrating planets
being swallowed up by the star. In giant stars there is also a lack of
planets with semi-major axis of $\leq$ 0.6 AU and possibly also a
different distribution of the eccentricities.  Sato et
al. (\cite{sato08}) discuss whether the lack of close-in planets in
giant stars is due to a difference in the formation of planets of more
massive star, or due to the fact that planets moves outwards when the
central stars loose mass, or whether the inner planets are engulfement
by their host stars when these expand.  The discovery of the planet of
the sdB star \object{V391 Peg} is very interesting in this respect as
this planet must have survived the red-giant phase, when the star
expanded to 70\% of the star-planet distance (Silvotti et
al. \cite{silvotti07}; Fortney \cite{fortney07}).  Once a sufficient
number of planets of giant and main-sequence stars that are more
massive than the sun are found, it will be possible to find out
whether the lack of planets with 0.6 AU in giant stars is due to the
formation, or evolution of the planets. If it were due to formation,
we would find the same lack of inner planets in main-sequence stars
more massive than the sun as in giant stars.

According to the theory by Kennedy \& Kenyon (\cite{kenedy08}), the
probability that a given star has at least one gas giant planet
increases linearly with stellar mass from 0.4 to 3 $\rm
M_\odot$. Kornet et al. (\cite{kornet06}) concluded that giant planets
tend to form in tighter orbits around {\em less} massive stars.
Although they find that the minimum metallicity at which planets can
form via core accretion decreases with increasing mass of the central
star, they also conclude that the frequency of massive planets should
be anticorrelated with the mass of the star.

Searching for planets around main-sequence stars more massive than the
sun is possible if one restricts the search to stars in the range
between 1.1 to about 1.7 $M_\odot$ (F-stars) and on the detection of
massive planets.  Such stars have still enough spectral lines and the
rotation rates are still modest compared to more massive stars so that
a reasonable RV precision is achieved.  Additionally, the mass of the
convection zone of F-stars is a factor 10 smaller than that of K stars
allowing us to investigate the atmospheric pollution hypothesis as
well (Pinsonneault, DePoy \& Coffee \cite{pinsonneault01}) .

In here we report on the indirect detection of a massive planet around
the F6V star \object{30 Ari B}, which is also part of a hierarchical
triple system.

\section{The host star 30 Ari B}

\object{30 Ari} is a visual binary star with a separation of
$38.2\pm0.7$ arcsec (Shatsky \cite{shatsky01}) consisting of the
components \object{30 Ari A} (HD 16246, HIP12189) and \object{30 Ari
  B} (HD 16232, HIP 12184). \object{30 Ari A} is a single line
spectroscopic binary.  Morbey \& Brosterhus (\cite{morbey74}) derive
an orbital period of $1.109526\pm0.000001$ days, an eccentricity of
$0.062\pm0.012$, and an amplitude of $22.41\pm0.31$ km\,s$^{-1}$.
Using the Hipparcos parallax of the two stars (Perryman et al. 
\cite{perryman97}, Zombeck \cite{zombeck06}) the distance to
\object{30 Ari A,B} is $39.8\pm0.3$~pc, which yields a projected
distance between the two stars of about $1520\pm54$ AU.  The orbital
period thus is likely to be larger than 10000 years.  As expected for
a close binary system, \object{30 Ari A} is an X-ray source and was
detected by ROSAT with an X-ray flux of $0.472\pm0.046$ ct/s (Zickgraf
et al. \cite{zickgraf03}). \object{30 Ari B} was not detected by
ROSAT.

\begin{figure}[h]
\includegraphics[width=0.45\textwidth, angle=0]{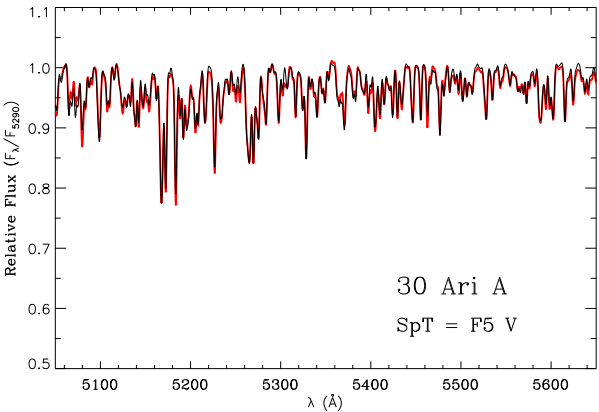}
\caption{Part of the TLS spectra of 30 Ari A. The observed spectra are
  displayed with thin lines, while the best fitting F5V templates are
  overplotted with thick lines. The spectra have been arbitrarily
  normalized to the flux at 5290 \AA.}
\label{SpecTypeA}
\end{figure}

\begin{figure}[h]
\includegraphics[width=0.45\textwidth, angle=0]{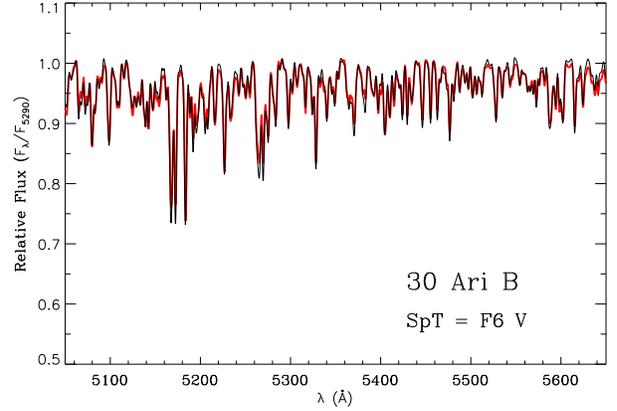}
\caption{Same as Fig.\,\ref{SpecTypeA} but for \object{30 Ari B}. 
  The best fitting template is an F6V star.}
\label{SpecTypeB}
\end{figure}

\object{30 Ari A} is listed in the SIMBAD database with a spectral
type F6III. If true, the object should have an absolute brightness
($\rm M_v$) of $+1.4$~mag. Using again the Hipparcos distance, the
relative magnitude ($m_v$) would than be $4.4$~mag, which is inconsistent
with the observed brightness (Tab.\,\ref{star}).  The V$-$J colours
are $0.808\pm0.02$ for \object{30 Ari A} and $1.02\pm0.02$ for
\object{30 Ari B}, and the V$-$K colours are $1.01\pm0.3$ and
$1.27\pm0.02$ for components A and B, respectively.  The colours are
also inconsistent with a giant star but consistent with a
main-sequence star. Using our high resolution spectra (see Sect.3) and
following the method described in Frasca et al. (\cite{frasca03}) and
Gandolfi et al. (\cite{gandolfi08}), we find that \object{30 Ari A} is
an F5V star and \object{30 Ari B}, and \object{30 Ari B} and F6V star
(Fig.\,\ref{SpecTypeA}; Fig.\,\ref{SpecTypeB}). Thus, the brightness,
colours and the results of the spectroscopy show that both components
are still on the main sequence stars.

\begin{figure}[h]
\includegraphics[width=0.35\textwidth, angle=-90]{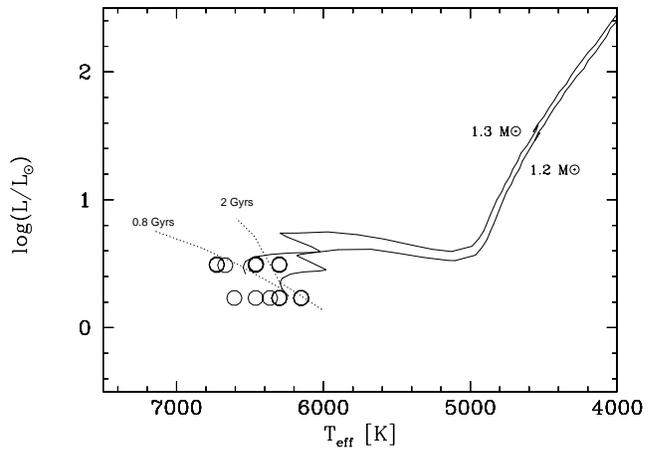}

\caption{The position of 30 Ari A and 30 Ari B in the
  Hertzsprung-Russell diagram, together with the evolutionary tracks
  from zero-age main-sequence stars to giant stars published by
  Girardi et al.  (\cite{girardi00}). The open circles show all values
  given in Tab.\,\ref{star} for 30 Ari A. The filled circles
  correspond to 30 Ari B. The tracks are for masses of 1.2 and
    1.3 $M_\odot$, and the isochrones for 0.8 and 2 Gyrs from that
    article. Using the method described in da Silva et
    al. (\cite{dasilva06}), we derive the parameters given in
    Tab.\,\ref{star}.}
\label{tracks}
\end{figure}

Fig.\,\ref{tracks} shows the position of both stars in the
Hertzsprung-Russell diagram, together with the evolutionary tracks
from Girardi et al. (\cite{girardi00}). Averaging the stellar
parameters and the using the method described in da Silva et
al. (\cite{dasilva06}), we derive the age, masses, and diameters of
the two stars (Tab.\,\ref{star}). We find that the mass of \object{30
  Ari A} and B are $1.31\pm0.04$ and $1.16\pm0.04$ $M_{\odot}$,
respectively.  For the ages of the two stars we derive $0.86\pm0.63$
and $0.91\pm0.83$ Gyrs.  The $v$\,sin\,$i$ values are also typical for
stars of that age and spectral type. The basic stellar data are
summarized in Tab.\,\ref{star}.  As can be seen from the various
abundance determinations both \object{30 Ari A} and \object{30 Ari B}
are slightly metal rich.

% A-component: HD 16246
% B-component: HD 16232

%%__________________________________________________________________
\begin{table}
\caption{Stellar properties of 30 Ari A,B }
\begin{tabular}{lll}
\hline \hline
Parameter                          & 30 Ari A                       & 30 Ari B                            \\
\hline
RA                                 &  02 37 00.5237                 &  02 36 57.7405                      \\
DEC                                & +24 38 50.0000                 & +24 38 53.0270                      \\

Spectral type                      &  F5V $^{14}$                  & F4V$^{4}$, F6V$^{6}$, F6V$^{14}$    \\
$\rm v\,sin\,i$ [$\rm km\,s^{-1}$] & $38.5\pm2.6^{14}$              & $40.6^{7}$, $38.3\pm1.8^{14}$       \\
V {\rm [mag]}                      & $6.497^{10}$, $6.48^{5}$       & $7.091^{10}$, $7.1^{5}$             \\
B-V {\rm [mag]}                    & $0.410^{8,13}$                 & $0.510^{12,13}$                     \\
J {\rm [mag]}                      & $5.681\pm0.019^{11}$           & $6.080\pm0.020^{11}$                \\
H {\rm [mag]}                      & $5.580\pm0.051^{11}$           & $5.908\pm0.029^{11}$                \\
K {\rm [mag]}                      & $5.479\pm0.024^{11}$           & $5.822\pm0.021^{11}$                \\
$\rm M_v$                          & $3.48^{10}$, $3.46^{5}$        & $4.12^{10}$, $4.12^{5}$             \\
$\pi$ {\rm [mas]}                  & $24.92\pm1.05^{3}$             & $25.36\pm1.10^{3}$                  \\
Distance {\rm [pc]}                & $40.1\pm1.7$                   & $39.4\pm1.7$                        \\
$\rm [Fe/H]$ {\rm [dex]}           & $0.27^{4}$, $0.00^{10}$,       & $0.27^{1,7}$, $-0.13^{10}$,         \\
                                   & $0.11^{2}$                     & $0.245\pm0.195^{9}$,                \\
                                   &                                & $0.03^{2}$, $0.27^{6}$              \\
$\rm T_{eff}$ {\rm [K]}            & $6462^{4}$, $6668^{10}$        & $6462^{1,4}$, $6152^{10}$           \\
                                   & $6300\pm60^{9}$,$6726^{2}$     & $6300\pm60^{9}$, $6364^{2}$,        \\
                                   & $6457^{5}$                     & $6607^{5}$                          \\
$\rm log\,g$                       & $4.51^{2}$, $4.25^{5}$         & $4.50^{4}$, $4.54^{2}$, $4.44^{5}$  \\
$\rm M_*\,[M_\odot]$               & $1.32\pm0.05^{10}$, $1.36^{5}$ & $1.11\pm0.06^{10}$, $1.20^{5}$      \\
                                   & $1.31\pm0.04^{14}$             & $1.16\pm0.04^{14}$                  \\
Age {\rm [Gyr]}                    & $0.86\pm0.63^{14}$             & $0.91\pm0.83^{14}$                  \\
$\rm R_*\,[R_\odot]$               & $1.37\pm0.03^{10}$             & $1.13\pm0.03^{14}$                  \\
$\rm \theta\,[mas]$                & $0.32\pm0.01^{14}$             & $0.26\pm0.01^{14}$                  \\
\hline\hline
\end{tabular}
\label{star}
\\
$^1$ Boesgaard \& Friel (\cite{boesgaard90}) \\
$^2$ Marsakov et al. (\cite{marsakov95})\\
$^3$ Hipparcos, Perryman and ESA (\cite {perryman97}) \\
$^4$ Cayrel de Strobel et al. (\cite{strobel97}) \\
$^5$ Allende Prieto et al. (\cite{allende99}) \\
$^6$ Malagnini et al. (\cite{malagnini00}) \\
$^7$ Buzzoni et al. (\cite{buzzoni01}) \\
$^8$ De Medeiros et al. (\cite{mediro02}) \\
$^9$ Taylor (\cite{taylor03}) \\
$^{10}$ Nordstr\"om et al. (\cite{nordstrom04}) \\
$^{11}$ Skrutskie et al. (\cite{skrutskie06}) \\
$^{12}$ Tolbert (\cite{tolbert64}) \\
$^{13}$ Mermilliod (\cite{mermilliod91})\\
$^{14}$ This work \\ 
\end{table}

\section{Observations}

\object{30 Ari B} was one of the stars monitored as part of the RV
planet search program of the Th\"uringer Landessternwarte as described
by Hatzes et al. (\cite{hatzes05}).  For this program we used the
2-m-Alfred Jensch telescope of the Th\"uringer Landessternwarte
Tautenburg which is equipped with an \'echelle spectrograph with
resolving power of $\rm \lambda/\Delta \lambda =67\,000$. During the
observations an iodine absorption cell was placed in the optical path
in front of the spectrograph slit. The resulting iodine absorption
spectrum that was superposed on top of the stellar spectrum provided a
stable wavelength reference for the measurement of the stellar RV. All
spectral observations were bias-subtracted, flat-fielded, and the
\'echelle orders extracted using standard IRAF routines.  

\section{Radial velocity measurements and orbit of the companion}

The RVs were calculated by modelling the observed spectra with a high
signal-to-noise ratio template of the star (without iodine) and a scan
of our iodine cell taken at very high resolution with the Fourier
Transform Spectrometer of the McMath-Pierce telescope at Kitt
Peak. The latter enables us to compute the relative velocity shift
between stellar and iodine absorption lines as well as to model the
temporal and spatial variations of the instrumental profile. See
Valenti et al. (\cite{valenti95}) and Butler et al. (\cite{butler96})
for a description of the principles behind this technique.  The median
of the errors of the RV measurements for our planet program stars is
about 9 m\,s$^{-1}$ and 11\% of these stars show RV variations of
$\leq 6$ m\,s$^{-1}$. This is based on the rms RV scatter for all
stars with trends or orbital solutions removed (when companions are
found). This error is mostly due to the S/N-ratio of the spectra and
not to the inherent accuracy of the instrument, since we can achieve a
precision of $\approx$ 3 m\,s$^{-1}$ on bright, slowly rotating
late-type stars. However, for F-stars the accuracy is much lower,
because of the fewer spectral lines.  Since \object{30 Ari B} is an
F-star and rapidly rotating ($v$\,sin\,$i$ $38.3\pm1.8$ $\rm
km\,s^{-1}$), and since average S/N-ratio of the spectra is only
$64\pm20$, it is not surprising that the errors of each individual
measurement is very large.  As usual the errors of the measurements
are derived from the standard-deviation of the RV values determined
for each of the $\sim$ 116 chunks into which the spectrum is divided
(Endl et al. \cite{endl04}; Hatzes et al.\cite{hatzes05}; Hatzes et
al.\cite{hatzes06}; D\"ollinger et al. \cite{doellinger07};
D\"ollinger et al. \cite{doellinger09}). The median RV error for
\object{30 Ari B} is $152$\,m\,s$^{-1}$. Table \ref{tab:RV} lists our
RV measurements for \object{30 Ari B}.

\begin{figure}[h]
\includegraphics[width=0.45\textwidth, angle=0]{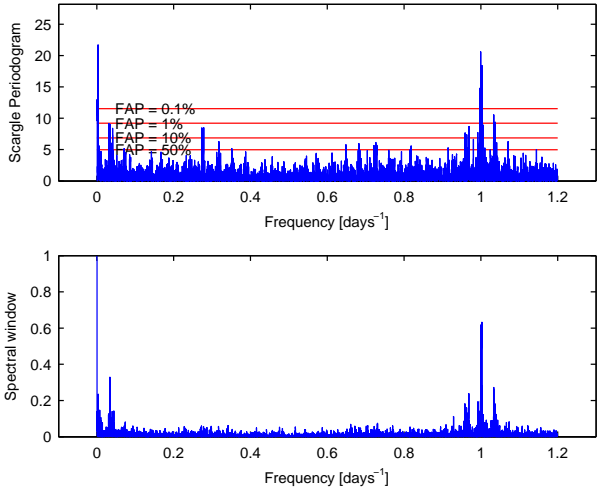}
\caption{Periodogram of the RV values. Apart from a one day alias,
there is only one peak at a frequency of 0.003 $\rm days^{-1}$. In the
upper panel, the false alarm probability is added.}
\label{periodRV}
\end{figure}

\begin{figure}[h]
\includegraphics[width=0.45\textwidth, angle=0]{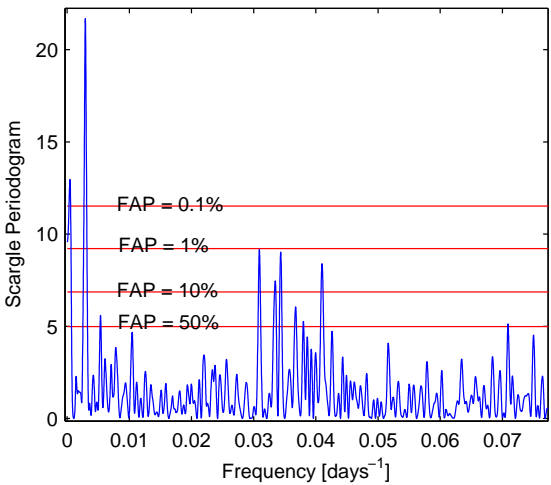}
\caption{Magnified periodogram close to the frequency 
of 0.003 $\rm days^{-1}$ (Fig.\,\ref{periodRV}) showing that
there is only one significant peak.}
\label{periodRV1}
\end{figure}

Fig.\,\ref{periodRV} shows a periodogram of the data obtained since
August 2002. There is only one significant peak at a frequency of
0.003 $\rm days^{-1}$. For clarity, we also show a magnified version
of this plot in Fig.\,\ref{periodRV1}. This peak is statistically
significant.  The false alarm probability (FAP) from the Lomb-Scargle
periodogram alone yields $\approx$ 10$^{-7}$. The FAP was also
estimated using a bootstrap randomization technique.  The measured RV
values were randomly shuffled keeping the observed times fixed and a
periodogram for each of these ``random'' data sets was then computed.
The fraction of the random periodograms having power higher than the
data periodogram yielded the false alarm probability that noise would
create the detected signal.  After 2$\times 10^{5}$ ``shuffles'' there
was no instance where the random data periodogram had power higher
than the real periodogram. This confirms the low value of the FAP.

\begin{figure}[h]
\includegraphics[width=0.50\textwidth,angle=0]{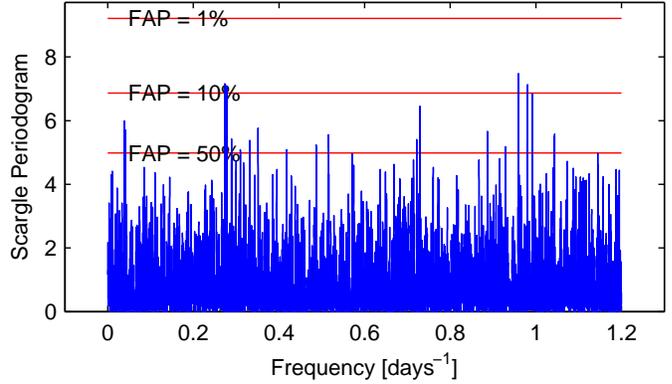}
\caption{Periodogram of the residuals of the RV after subtracting
the orbit: There is no indication for another companion.}
\label{periodRVres}
\end{figure}

The orbital solution was determined using the non-linear least squares
fitting program {\it Gaussfit} (Jefferys et al. \cite{jefferys88}).
These orbital parameters are listed in Table\,\ref{tab:orbit}.  
Fig.\,\ref{periodRVres} shows the periodogram of the RV residuals
after subtracting the orbital solution. There is no indication of
additional signals (companions) in the RV data.

\begin{figure}[h]
\includegraphics[width=0.45\textwidth, angle=0.0]{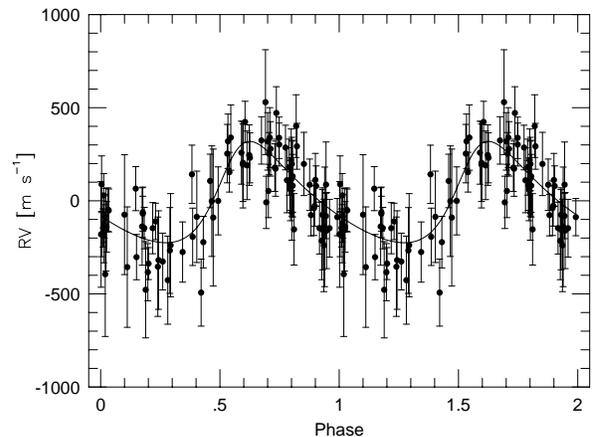}
\caption{Phase-folded RV curve together with the orbit.}
\label{orbit}
\end{figure}

\begin{figure}[h]
\includegraphics[width=0.35\textwidth, angle=270]{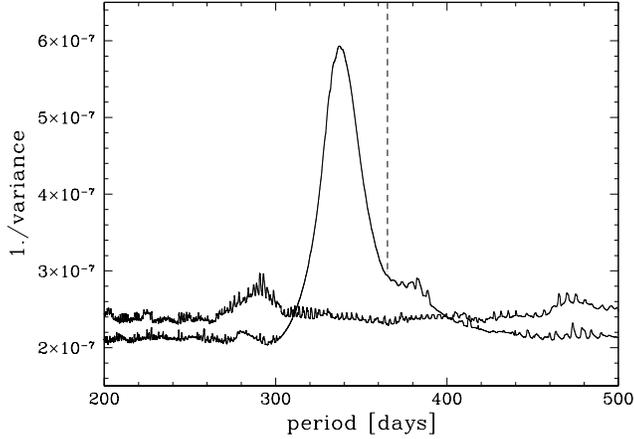}

\caption{Shown are the $\rm 1/variance$ of residuals between the RV
  measurements and the best-fitting Kepler orbit for a given
  period. The peak indicates that the difference between a Kepler
  orbit and the measurements is minimal for a period of 335 days,
  clearly different from one year (vertical dashed line).  Also shown 
  is $\rm 1/variance$ for just one  of the permutated data sets
  demonstrating that there is no alias close to the 335-day period.}
\label{periodRVlarge}
\end{figure}

\begin{figure}[h]
\includegraphics[width=0.35\textwidth,angle=270]{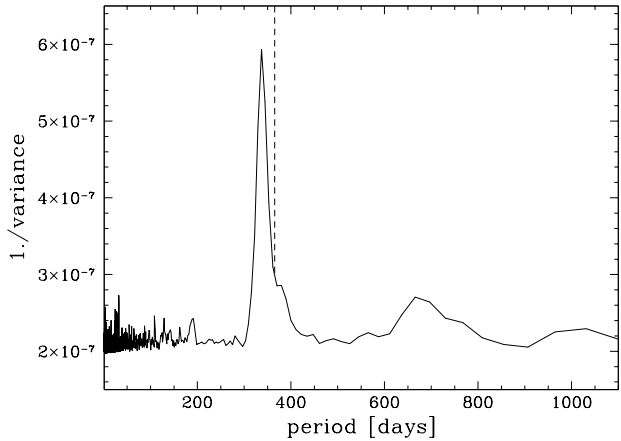}

\caption{Same as Fig.\,\ref{periodRVlarge} but for the period range
  between 1.0 and 1100 days demonstrating that the largest peak is at
  a period of 335 days.  }
\label{periodRVlong}
\end{figure}

The classical Lomb-Scargle periodogram analysis is less than ideal for
detecting planets in eccentric orbits. A better way to detect these is
to phase-fold the data to all possible periods within a certain range
and then fitting a Kepler orbit to the data for each period.  As a
test we also fitted all possible Kepler orbits with periods between
less than one day and 2213 days, the time for which we monitored the
object. We also varied the eccentricity between 0 and 0.9.  The best
orbital solution is the one that minimizes the variance between it and
the RV measurements.  Thus, $\rm 1/variance$ is maximized and this can
be used as a measure of the quality of the fit. A high value of $\rm
1/variance$ indicates a small difference between the best Kepler orbit
and the data. As can easily be seen in Fig.\,\ref{periodRVlarge} and
Fig.\,\ref{periodRVlong} the Kepler orbit which matches best the
observed data has a period of 335 days. The parameters of the
corresponding orbit are consistent with the {\em Gaussfit} solution,
which is shown as a line in Fig.\,\ref{orbit}.  The values derived
given in Table\,\ref{tab:orbit}.  With a stellar mass of $1.16\pm0.04$
$\rm M_\odot$, the minimum companion mass is $9.88\pm0.94$ $\rm
M_{Jup}t$.

\begin{figure}[h]
\includegraphics[width=0.45\textwidth, angle=0.0]{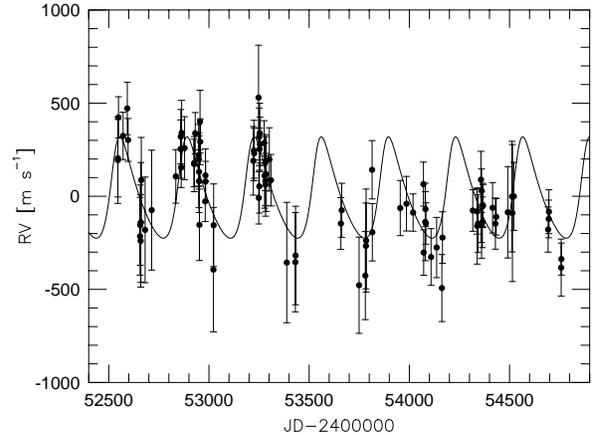}
\caption{RV measurements of 30 Ari B. The solid line is the orbital
  solution.  }
\label{orbitHJD}
\end{figure}

Unfortunately, the orbital period is close to one year which makes it
difficult to cover all phases. Because of the relatively high latitude
of Tautenburg observatory (+51 degrees) managed, however to obtain RV
measurements of consecutive minima and maxima of the RV curve
(Fig.\,\ref{orbitHJD}).  Further tests are nevertheless required in
order to find out whether this period is in fact a one year alias, or
not. To test this hypothesis we permutated randomly the order of the
RV measurements in respect to the time when the observations were
taken, and applied the same period finding algorithm as before. If the
period were due to an alias we should find the same period in the
permutated as in the original data. The second line in
Fig.\,\ref{periodRVlarge} shows the periodogram just for one such
permutated data set.  We repeated the same experiment with 2$\times
10^{5}$ different permutations and none of these showed any
significant peak in that period range.

Furthermore, the orbital period differs by more than 5 sigma from a
1-year period (Fig.\,\ref{periodRVlarge}).  This excludes that the
signal is an artifact of incomplete removal of the earth's barycentric
motion. We thus conclude that the 335-day period in the RVs real.
Fig.\,\ref{orbitHJD} shows the RV measurements together with the orbit
and Fig.\,\ref{orbit} the phase-folded RV measurements.

\section{Activity or companion?}

We find RV variations of \object{30 Ari B} with a period of
$335.1\pm2.5$ days. What is the nature of the RV variations? It is
unlikely that this period is due to oscillations as such long-period
pulsations have never been found in main-sequence stars.  From the
radius of $1.13\pm0.03$ $\rm R_\odot$, and the $v$ sin $i$ of the star
we derive that the rotation period has to be $\leq 1.5$ days. It is
thus impossible that 335 days is the rotation period of the star. In
principle, another possibility would be that the RV variations are
caused by an activity cycle.  Variations on long time scales might be
caused by changes of the granulation, which not only effects the RV
but also the central depth and thus the equivalent width of
photospheric lines (Livingston et al. \cite{livingston07}).

According to Ossendrijver (\cite{ossendrijver96}) the length of the
activity cycle is given by the relation: $P_{cyc}=
6.9\,P^{2\pm0.3}_{rot}\,\tau^{-2\pm0.3}_c$ [yrs], with the convective
turnover time $\tau_c$.  The convective turnover time would thus have
to be of the order of four days in order to explain an activity cycle
$P_{cyc}$ of $0.92$ years. A $\tau_c$ of four days is certainly not
very plausible for an F6V star (Ossendrijver \cite{ossendrijver96}).

\begin{figure}[h]
\includegraphics[width=0.50\textwidth, angle=0]{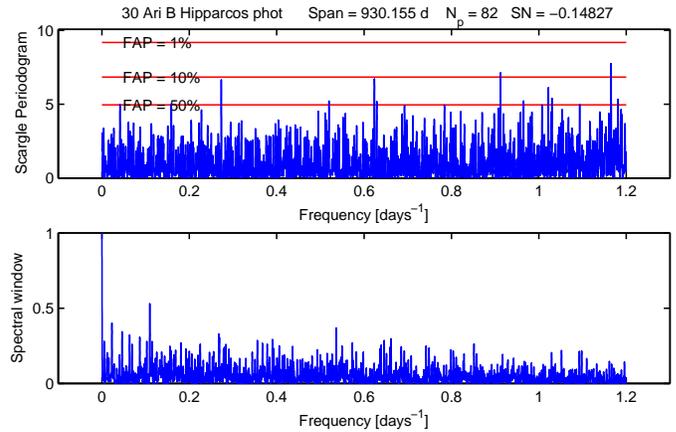}
\caption{Periodogram of the Hipparcos photometry. There is
no peak at a frequency of 0.003 $\rm days^{-1}$.}
\label{periodPhot}
\end{figure}

\begin{figure}[h]
\includegraphics[width=0.35\textwidth, angle=270]{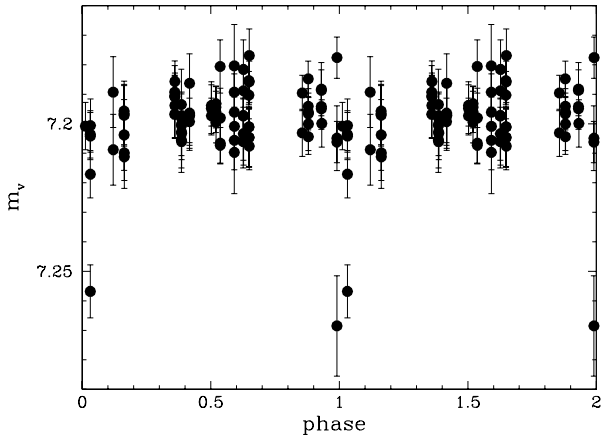}
\caption{Shown are photometric measurements taken with the Hipparcos 
satellite of \object{30 Ari B}, phase-folded to the same period as
in  Fig.\,\ref{orbit}. The brightness of the star does not vary in
phase with the RV.}
\label{phot}
\end{figure}

From the relationship between the filling factor and the RV jitter
from Saar \& Donahue (\cite{saar97}), we derive that the brightness of
the star would have to change by $\geq 1$\% if the observed RV
variations were caused by spots.  In order to test this hypothesis, we
examined the photometric measurements taken with the Hipparcos
satellite (Perryman and ESA \cite{perryman97}). Fig.\,\ref{periodPhot}
shows the periodogram of the Hipparcos photometry and Fig.\,\ref{phot}
the values phase-folded to the orbital period given in
Table\,\ref{tab:orbit}.  The photometric measurements do not show any
periodicity associated with the planet period. We should note,
however, that the Hipparcos photometry was not contemporaneous with
our RV measurements. The average brightness of star in the interval of
$\pm$ 0.1 in phase around the maximum and minimum of the RV is
$7.1964\pm0.0015$, and $7.1969\pm0.0015$ mag, respectively.  Thus, the
star has the same brightness at both phases which implies that there
no correlation between the brightness of the star and the RV.

%%__________________________________________________________________
\begin{table}
\caption{Orbital elements of 30 Ari B}
\begin{tabular}{ll}
\hline \hline
Element & Value \\
\hline
Period [days]                   & $335.1\pm2.5$ \\
$T_0$ [HJD]                     & $245\,4538\pm20$ \\
$K_{1}$ [$\rm m\,s^{-1}$]        & $272\pm24$ \\
$e$                             & $0.289\pm0.092$ \\
$\omega$ [deg]                  & $307\pm18$  \\
$\sigma $ (O-C) [$\rm m\,s^{-1}$]  & 135 \\
$f(m)={{m_2^3 \sin^3{i}}\over{(m_1+m_2)^2}}$ [$\rm M_\odot$] & $(6.1\pm1.7)\,10^{-7}$ \\
$a_1$\,sin\,$i$ [AU]              & $0.00802\pm0.00074$ \\
$a$ [AU]                        & $0.995\pm0.012$ \\
$m_1$ [$\rm M_\odot$]            & $1.16\pm0.04$  \\
$m_2$\,sin\,$i$ [$\rm M_{Jup}$]    & $9.88\pm0.94$ \\
\hline\hline
\end{tabular}
\label{tab:orbit}
\end{table}

\begin{figure}[h]
\includegraphics[width=0.35\textwidth, angle=270]{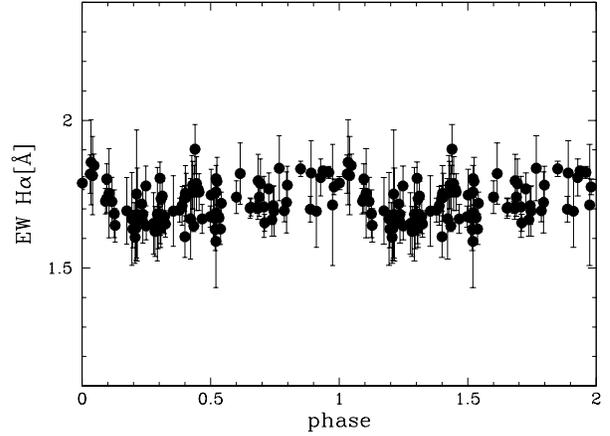}
\caption{Shown are measurements of the equivalent with
of H$\alpha$ phase-folded to the same period as
in  Fig.\,\ref{orbit}. The equivalent width does not vary in
phase with the RV.}
\label{Halpha}
\end{figure}

As an additional test, we analyzed the strength of the H$\alpha$ and
H$\beta$ lines.  If we would find that the equivalent width, or the
depth of H$\alpha$ and H$\beta$ were correlated with the RV, we would
have to conclude that the RV variations are caused by stellar activity
(K\"onig et al. \cite{koenig05}).  Fig.\,\ref{Halpha} shows the
equivalent width measurements of H$\alpha$ phase-folded in the same
way as Fig.\,\ref{orbit}. Even after binning the data in phase, we do
not see any obvious variations that are in phase with the RV
measurements. The equivalent width of H$\alpha$ variability is not
correlated with the RV.

\begin{figure}[h]
\includegraphics[width=0.35\textwidth, angle=270]{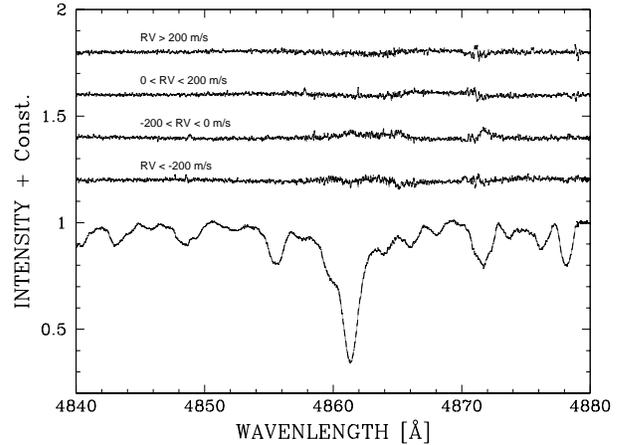}

\caption{Average spectrum of \object{30 Ari B} in the H$\beta$ region.
  The four upper curves show the ratio of the averaged spectra in the
  four RV intervals ($RV<-200\,m/s$, $-200<RV<0\,m/s$,
  $0<RV<200\,m/s$, and $RV>200\,m/s$) to the average spectrum in this
  region. There is no indication for a correlation between the RV and
  the depth of H$\beta$.}

\label{Hbeta}
\end{figure}

As another test, we averaged all spectra in the RV intervals $\rm
RV<-200$, $\rm -200<RV<0$, $\rm 0<RV<200$, and $\rm RV>200m\,s^{-1}$,
and then divided these four spectra by the average of all spectra.
The result is shown in Fig.\,\ref{Hbeta} for the spectral region
containing H$\beta$. Again, there is no evidence that the depth of
H$\beta$ is correlated with the RV.

\begin{figure}[h]
\includegraphics[width=0.35\textwidth, angle=270]{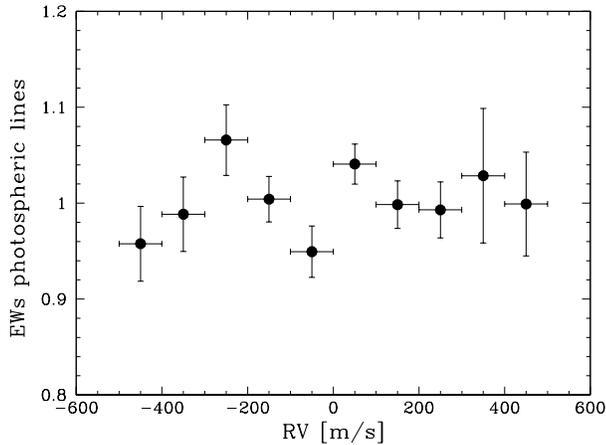}
\caption{Shown are measurements of the average equivalent with of
  photospheric lines against the RV.  There is no correlation
between RV and the equivalent with of the photospheric lines.}
\label{photlines}
\end{figure}

As a final test we did the same analysis for the photospheric
lines. Unfortunately, the number of photospheric lines that are not
effected by the iodine lines are rather limited.
Fig.\,\ref{photlines} shows the averaged equivalent width of
photospheric lines that are not effected by the iodine versus the
RV. As before, we binned the data into velocity intervals. Again,
there is no correlation between the equivalent width and the radial
velocity. Since all four tests show no indication that the RV
variations are caused by stellar activity, we conclude that they are
caused by a companion.

\section{Discussion and conclusions}

We took RV measurements of the F-star \object{30 Ari B} over a time
span of six years. The RV shows periodic variations with a period of
$335.1\pm2.5$ days, which can be fitted with a Keplerian orbit with an
eccentricity of $0.289\pm0.092$. Since the period remained unchanged
during this time, and since the star shows neither significant
photometric variations, nor significant variations of the Balmer, or
photospheric lines, the best interpretation is that the RV variations
are caused by an orbiting body.

This star is part of a hierarchical triple system.  However, the large
separation between \object{30 Ari A} and \object{30 Ari B} (about 1500
AU) is much larger than that of $\gamma$ Cep (Hatzes et
al. \cite{hatzes03}) which has a binary separation of about 40 AU. A
detailed study by Desidera \& Barbieri (\cite{desidera07}) shows that
the properties of planets of wide binaries (projected separation of
about 200-300 AU) and orbital periods larger than 40 days are the same
as that of single stars.  Thus, the binary nature of \object{30 Ari A}
is presumably unimportant for \object{30 Ari B\,b}.

We have measured a rotational velocity of 38.3 $\pm$ 1.8 km\,s$^{-1}$
for \object{30 Ari B}.  This rotational rate implies that we most
likely are viewing the star nearly equator-on.  Nordstr\"om et
al. (\cite{nordstrom97}) measured rotational velocities of 592 early
F-type stars. The median rotation rate was about 50 km\,s$^{-1}$. The
average effective temperature for their sample was 6860 K which means
that \object{30 Ari B} has a slightly later spectral type than a
typical member of the Nordstr\"om et al. (\cite{nordstrom97}) sample.
Assuming that the orbital and stellar spin axes are aligned, sin\,$i$
is most likely near unity.  The minimum mass derived thus is probably
close to the true mass.  

In any case a minimum mass of $9.88\pm0.94$ $\rm M_{Jup}$ implies that
the true mass is likely to be close to the planet/brown dwarf
boundary. As pointed out by Udry \& Santos (\cite{udry07}), while most
gaseous planets have masses below 5 $M_{Jup}$, the distribution has a
long tail. A statistical analysis of all available data by Grether \&
Lineweaver (\cite{grether06}) shows that the driest part of the brown
dwarf desert is at $M=31^{25}_{-18}\,M_{Jup}$. These authors also find
that $11\pm3\%$ of the solar-like stars are binaries, $5\pm2\%$ have
giant planets but less than $1\%$ have close brown dwarf companions
with orbital periods of less than 5 years.  Given the fact that the
inclination for most of these systems is not known, some of the
suspected brown dwarf companions may even turn out to be binary
stars. The only transiting object with a mass in the brown dwarf
regime orbiting a star is CoRoT-Exo-3b ($M=21.7\pm1.0\,M_{Jup}$)
(Deleuil et al. \cite{deleuil08}). Such objects thus are rare but they
do exist.

The lack of high-mass planets is particularly striking for planets
with an orbital period of less than 100 days. In fact, Udry \& Santos
(\cite{udry07}) list only three planets where the mass is higher, and
the orbital period is shorter than that of \object{30\,Ari\,B\,b}.
According to Eggenberger et al. (\cite{eggenberger04}) most of the
very massive planets are in binary systems. But as pointed out above,
the distance to \object{30 Ari A} is presumably too large to
have any effect on the properties of the planet.

If we compare \object{30 Ari B\,b} with other extrasolar planets, then
its closest match is \object{HD 33564\,b} (Galland et al.
\cite{galland05b}): The host star also is an F6V star, $m$\,sin\,$i$ =
9.1 $\rm M_{Jup}$, $P$\,=\,388\,days, and $e$ = 0.34.  \object{30 Ari
  B\,b} thus might belong to a class of objects that is not exotic but
simply represents the extension of the distribution of planets into
the brown dwarf regime.

Although \object{30 Ari B\,b} is the ninth planet discovered by the
Tautenburg survey around a star with a mass higher than the sun, a
statistically larger sample is needed before concluding that very
massive planets are more common among stars more massive than the sun.

%%__________________________________________________________________
\begin{table}
\caption{RV measurements}
\begin{tabular}{cr}
\hline \hline
HJD-2450000 & RV [$m\,s^{-1}$]\\
\hline 
2545.536478  &  194  $\pm$  232 \\
2545.544754  &  204  $\pm$  208 \\
2548.472892  &  424  $\pm$  111 \\
2571.576275  &  324  $\pm$  127 \\
2592.467015  &  471  $\pm$  141 \\
2596.415295  &  302  $\pm$  115 \\
2656.242768  &  -215  $\pm$ 209 \\
2657.267185  &  -155  $\pm$ 215 \\
2659.260606  &  -240  $\pm$ 248 \\
2660.294048  &  -140  $\pm$ 321 \\
2662.268544  &  87  $\pm$  229 \\
2681.341841  &  -181  $\pm$ 284 \\
2714.276746  &  -75  $\pm$  322 \\
2834.523535  &  107  $\pm$  144 \\
2858.575149  &  253  $\pm$  157 \\
2859.600468  &  320  $\pm$  146 \\
2861.586390  &  154  $\pm$  107 \\
2863.588342  &  340  $\pm$  175 \\
2878.477610  &  258  $\pm$  169 \\
2925.422421  &  178  $\pm$  128 \\
2926.480403  &  172  $\pm$  148 \\
2931.400890  &  338  $\pm$  111 \\
2948.365052  &  219  $\pm$  167 \\
2949.384915  &  198  $\pm$  143 \\
2950.438364  &  132  $\pm$  193 \\
2950.444684  &  81  $\pm$  168 \\
2952.400062  &  -153  $\pm$  192 \\
2955.422149  &  401  $\pm$  168 \\
2956.465067  &  293  $\pm$  123 \\
2981.367219  &  -27  $\pm$  109 \\
2982.388335  &  112  $\pm$  143 \\
2983.413199  &  79  $\pm$  108 \\
3022.290291  &  -394  $\pm$ 334 \\
3023.367630  &  -156  $\pm$ 222 \\
3221.496697  &  190  $\pm$  184 \\
3224.547386  &  245  $\pm$  162 \\
3225.512468  &  231  $\pm$  134 \\
3247.392319  &  530  $\pm$  281 \\
3248.490970  &  -8  $\pm$  140 \\
3250.480289  &  319  $\pm$  154 \\
3251.466370  &  53  $\pm$  143 \\
3252.492894  &  253  $\pm$  114 \\
3253.432338  &  339  $\pm$  161 \\
3254.450110  &  279  $\pm$  146 \\
3275.513744  &  286  $\pm$  120 \\
3277.546322  &  110  $\pm$  175 \\
3280.508460  &  181  $\pm$  150 \\
3281.621507  &  83  $\pm$  134 \\
3282.493267  &  62  $\pm$  169 \\
3284.350643  &  119  $\pm$  201 \\ 
3301.371752  &  198  $\pm$  170 \\
3309.444008  &  86  $\pm$  138 \\
3388.344614  &  -356  $\pm$  323 \\
3431.291948  &  -354  $\pm$  267 \\
3432.265935  &  -318  $\pm$  265 \\
3658.364855  &  -147  $\pm$  139 \\
3662.535227  &  -75  $\pm$  145 \\
3749.258017  &  -478  $\pm$  258 \\
3780.366708  &  -426  $\pm$  236 \\
3783.265338  &  -267  $\pm$  230 \\
3784.288850  &  -238  $\pm$  277 \\
3814.276631  &  142  $\pm$   157 \\
3815.290494  &  -193  $\pm$  154 \\
3954.589497  &  -63  $\pm$  148 \\
3985.625596  &  -40  $\pm$  147 \\
4018.454970  &  -88  $\pm$  100 \\
4070.394848  &  65  $\pm$  118 \\
4071.515854  &  -302  $\pm$  122 \\
4079.374428  &  -138  $\pm$  208 \\
4080.365276  &  -69  $\pm$  94 \\
4082.438154  &  -146  $\pm$  111 \\
4108.340755  &  -326  $\pm$  152 \\
4136.254291  &  -275  $\pm$  162 \\
4162.258663  &  -493  $\pm$  180 \\
4165.348658  &  -222  $\pm$  138 \\
4316.580289  &  -77  $\pm$  111 \\
4337.564193  &  -82  $\pm$  122 \\
4338.559366  &  -160  $\pm$  204 \\
4342.615233  &  -147  $\pm$  155 \\
4357.560127  &  89  $\pm$  152 \\
4359.572735  &  -146  $\pm$  102 \\
4360.541338  &  -153  $\pm$  180 \\
4360.576121  &  29  $\pm$  117 \\
4364.560715  &  -140  $\pm$  136 \\
4366.536859  &  -47  $\pm$  117 \\
4367.554451  &  -52  $\pm$  115 \\
4415.427584  &  -62  $\pm$  136 \\
4429.384583  &  -147  $\pm$  137 \\
4433.367364  &  -111  $\pm$  98 \\
4491.292838  &  -86  $\pm$  245 \\
4512.358374  &  -2  $\pm$  297 \\
4514.353088  &  -90  $\pm$ 368 \\
4521.285415  &  0  $\pm$  183 \\
4692.598912  &  -179  $\pm$ 121 \\
4695.601537  &  -122  $\pm$  101 \\
4696.539080  &  -83  $\pm$  118 \\
4757.636700  &  -383  $\pm$  153 \\
4758.610225  &  -337  $\pm$  86 \\
\hline\hline 
\end{tabular}
\label{tab:RV}
\end{table}

% -----------------------------------------------------------------------
\begin{acknowledgements}

We are grateful to the user support group of the Alfred-Jensch
telescope. APH and MH acknowledge the support from the Deutsche
Forschungsgemeinschaft (DFG) of grant HA 3279/3-2, EWG and FC of grant
GU 464/11-1.  ME acknowledges a grant by the DFG-Graduiertenkolleg
1351 ``Extrasolar Planets and their Host Stars''. Additionally, DG
acknowledges the support of a grant from the Deutschen Zentrums f\"ur
Luft- und Raumfahrt (DLR) (50OW0204).  This research has made use of
the SIMBAD database, operated at CDS, Strasbourg, France.  This
publication makes use of data products from the Two Micron All Sky
Survey, which is a joint project of the University of Massachusetts
and the Infrared Processing and Analysis Center/California Institute
of Technology, funded by the National Aeronautics and Space
Administration and the National Science Foundation.

\end{acknowledgements}

\end{document}